\PassOptionsToPackage{unicode}{hyperref}
\PassOptionsToPackage{hyphens}{url}
\documentclass[reprint,amsmath,amssymb,aps]{revtex4-2}
\usepackage{lmodern}
\usepackage{xcolor}
\usepackage[margin=1in]{geometry}
\usepackage{longtable,booktabs}
\usepackage{etoolbox}
\usepackage{hyperref}

\usepackage[T1]{fontenc}
\usepackage[utf8]{inputenc}
\usepackage{textcomp} 

\usepackage{graphicx,grffile}
\makeatletter
\def\maxwidth{\ifdim\Gin@nat@width>\linewidth\linewidth\else\Gin@nat@width\fi}
\def\maxheight{\ifdim\Gin@nat@height>\textheight\textheight\else\Gin@nat@height\fi}
\makeatother
\setkeys{Gin}{width=\maxwidth,height=\maxheight,keepaspectratio}
\makeatletter
\def\fps@figure{htbp}
\makeatother
\begin{document}

\title{Surrogate Monte Carlo}
\author{A. Christian Silva}
\email{csilva@idatafactory.com}
\affiliation{
 www.idatafactory.com\\
 Stuart, FL, USA
}
\author{Fernando F. Ferreira}
\email{ferfff@usp.br}
\affiliation{
 Department of Physics-FFCLRP\\ 
 Universidade de S\~ao Paulo (USP) \\ 
 Ribeirao Preto-SP, 14040-901, Brazil \\
 and \\
 Centre for Interdisciplinary Research on Complex Systems \\
 Universidade de S\~ao Paulo (USP) \\
 03828-000 S\~ao Paulo, Brazil
}

\begin{abstract}
This article proposes an artificial data generating algorithm that is simple and easy to customize. The fundamental concept is to perform random permutation of Monte Carlo generated random numbers which conform to the unconditional probability distribution of the original real time series. Similar to constraint surrogate methods, random permutations are only accepted if a given objective function is minimized. The objective function is selected in order to describe the most important features of the stochastic process. The algorithm is demonstrated by producing simulated log-returns of the S\&P 500 stock index.
\end{abstract}

\maketitle

\hypertarget{introduction}{%
\section{Introduction}\label{introduction}}

The generation of artificial time series has been traditionally associated with Monte Carlo simulations. For instance, one assumes a stochastic process and finds the parameters of such process by fitting the stochastic model to the observed time series. Once the parameters are known, the researcher generates artificial data by Monte Carlo simulations. One of the most simple examples is the use of Geometric Brownian motion to model stock prices \cite{glasserman2004monte}. Other examples span the most diverse research areas from fluid dynamics to trafic flow \cite{Grassia1995,McKane2003,Schweitzer2003}.

In recent years, deep learning methods have proposed an alternative to the traditional Monte Carlo simulation \cite{Esteban2017,GA2020}. The idea of using deep neural networks can bypass the specification of the stochastic process and therefore be model-free. In this scenario the researcher has to train the model to the  empirical time series without necessarily understanding the underlying dynamics. The benefit of such methodology is that one can, in principle, produce high fidelity artificial data with features that can not be modeled by traditional methods either due to the complexity or because prior research was not able to uncover such features. The risk of using deep learning is of course overfiting and, for some, the black-box nature of such models.

This article proposes a methodology that is also model-free by combining random number generation and surrogates. Surrogates have a long history in the physical sciences most notably in the non-linear dynamics/chaos communities \cite{kantz2004nonlinear}. The general idea is to transform the original time series by performing suitable permutations of the observations such that the resulting time series is composed by exactly the same observations but in a different order. This order is chosen to satisfy some constraints (constraint surrogate generation method by \cite{Schreiber1998}). The goal is to perform hypothesis tests by comparing the original time series and the surrogate since both have exactly the same unconditional distribution \cite{LANCASTER20181}. A typical application is to test whether the original time series is generated by a non-linear process by comparing surrogates which preserve the linear auto-correlations (power spectrum) of the original time series but nothing else \cite{LANCASTER20181}. If one requires more and more constraints on the permutations when generating surrogates less surrogates can be generated. In the limit, if one insists on having dynamics which are identical to the original time series, the only surrogate is the time series itself. The recipe for producing surrogate time series which are statistically similar to the original is to require the ``correct'' set of constraints which capture the essence of the dynamics. 

Surrogates traditionally are created using the exact same observations of the measured time series which is not desired when simulating. The idea is to simulate from the ``correct'' joint probability distribution without necessarily modeling such distribution. The method proposed here is to first draw random numbers of the unconditional stationary probability distribution of the original time series. This creates an independent and identically distributed time series of random numbers. One then permutes such random numbers imposing constraints much like what is done for surrogates \cite{Schreiber1998}. This procedure produces an artificial time series with joint probability distribution that agrees to the measured joint probability density within some accuracy.

Notice, however, that this procedure might not converge since contrary to the standard surrogate method, we are now applying permutations of random numbers which have a low probability of samples which are corner cases (to many zeros etc). If that happens no permutation might satisfy the constraints within the desired accuracy. This risk is particularly high if the time series is ``small'' and the number of constraints is ``high''.

The next sections detail the algorithm and illustrate the methodology by comparing artificial daily log-returns of the S\&P 500 with the actual returns. Finally, the appendix presents further details on the convergence of the algorithm as well as surrogate Monte Carlo applied to a known AR(1) model as a simple sanity check.

\hypertarget{unconditional-probability-density}{%
\section{Unconditional probability density}\label{unconditional-probability-density}}

There are many strategies that can be used to produce random numbers with an unconditional probability distribution that agrees approximately with observations. Parametrically, one popular approach in finance is to assume student's t-distribution \cite{LB2018}. However, the choice here is to remain non-parametric and to adapt the inverse transform method \cite{glasserman2004monte} to the discrete empirical distribution of the asset returns.

The algorithm is as follows. First one builds the empirical cumulative distribution function (CDF) of the asset returns \(X\). In practice, the CDF is a table of values that maps a given asset return (\(x\)) to a number between zero and one (\(u\)). The next step is to generate uniform random numbers between zero and one (\(U \sim Unif[0,1]\)). Finally the algorithm finds the asset return \(x\) which corresponds to the number \(u\) by looking at the CDF table.

Note that both \(U\) and \(X\) are real numbers and therefore there is no perfect match to any value in the table. This issue is resolved by linear interpolation. Therefore any value of \(u\) will have a corresponding \(x\) except when \(u\to 0,1\). These limiting cases are extrapolations which have to be decided by additional considerations.

In this work, the choice is to truncate to the largest and smallest historical \(x\). Therefore when \(u\to 0,1\) the algorithm takes \(x \to x_{min},x_{max}\). This is the most simple option, however a parametric model for the tails could be a better choice since it is very difficult to envision any other method given the very few data points.

\begin{figure}
\centering
\includegraphics{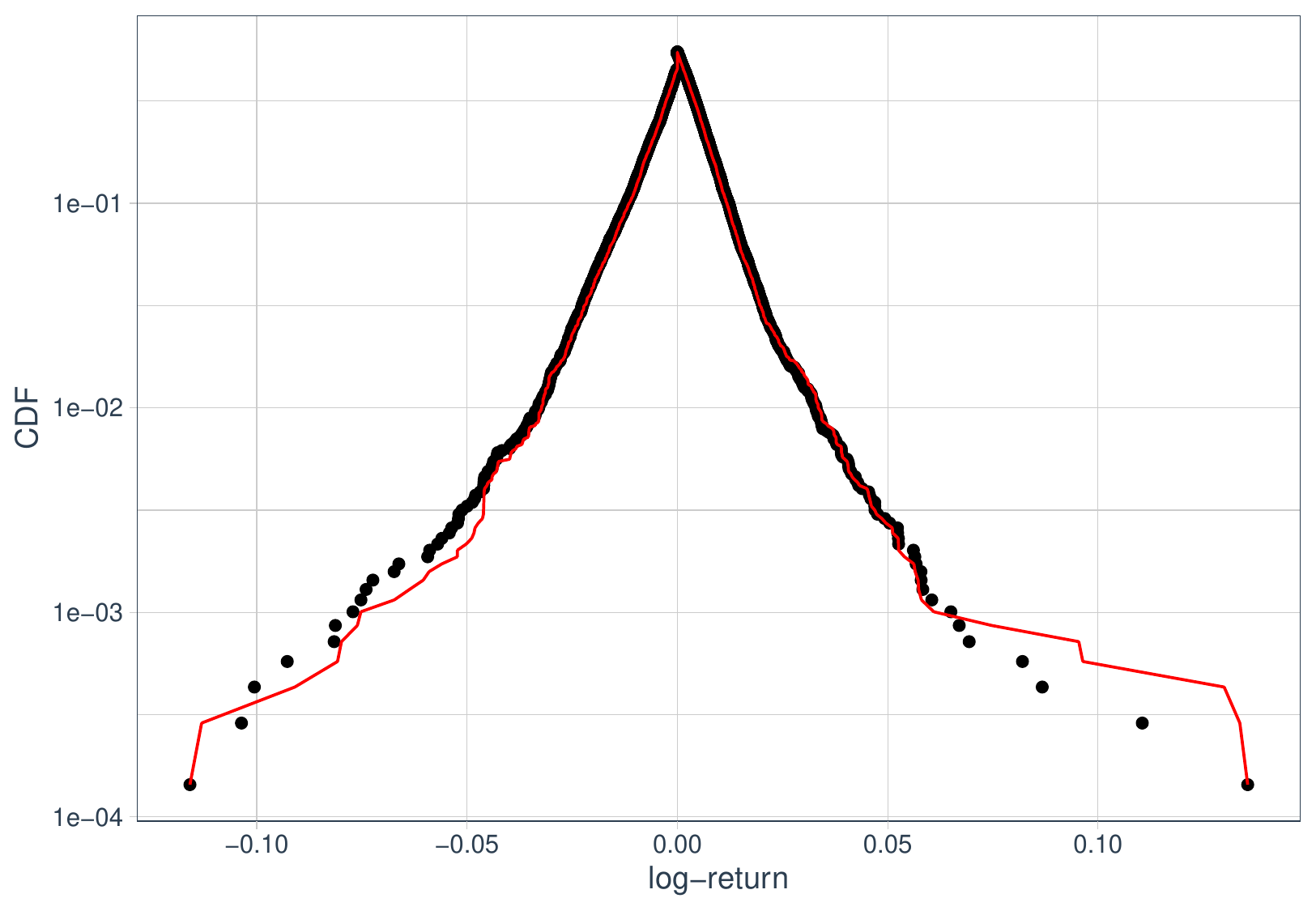}
\caption{\label{fig:cdfFig}Cumulative distribution function (CDF) of the S\&P 500 daily log-returns from 1993 to 2020 (black circles) together with the monte carlo generated log-returns (red solid line).}
\end{figure}

Figure \ref{fig:cdfFig} compares the empirical CDF of the daily S\&P 500 log-returns with the Monte Carlo generated log-returns of price \(P\) defined by

\begin{equation}
x_t =  \log{\frac{P_{D}}{P_{D-t}}} \label{eq:log1}
\end{equation}

where \(x_t\) is the return over time interval \(t\) and \(P_{D}\) is the price on day \(D\). The SPY exchange traded fund is used as a proxy for the S\&P 500 prices with data adjusted for dividends. The close prices are downloaded from Yahoo finance go from 1993-02 to 2020-08, a total of 6930 days.

Figure \ref{fig:cdfFig} shows that the agreement between data and simulated data is good over many orders of magnitude. Notice that the CDF is folded (also know as mountain plot \cite{Katherine1995}), that is, if the \(CDF(x)\) is defined as \(\int_{-\infty}^x p(u) du\) then the plot shows \(CDF(x)\) for \(x<=0\) and \(1-CDF(x)\) for \(x>0\). This particular presentation of the CDF highlights the tails of the distribution which are much harder to estimate and simulate.

\hypertarget{surrogates}{%
\section{Surrogates}\label{surrogates}}

The idea of using the constraint randomization algorithm introduced by \cite{Schreiber1998} precludes that one has a set of features which can be measured and that these features are enough in order to approximate the joint probability density of which the measured time series is one realization.

The key issue is to select such features and it is here that deep learning methods are most practical since in principle these methods would not require the end user to know such features a priori. This automation is not without dangers and represents a paradigm shift which is sometimes at odds with the traditional scientific method of recognizing the fundamental dynamical drivers \cite{efron2020}. The current work takes a more traditional approach, however, learning algorithms could be used in conjunction with the ideas proposed here (to identify such features for instance).

The approach taken here leverages extensive research validated in the past 25 years that has identified some of the most relevant features for financial assets (stylized facts) \cite{Chakraborti2011}. In particular volatility clustering and the leverage effect are well documented features which should be present when producing artificial asset returns. The first is responsible for the long memory dependence of the volatility and can be identified by the slow decay of the auto-correlation of the absolute returns and the last by the relation between future volatility and present returns.

Given \(N\) observed elements \(x_t\) and a Monte Carlo generated time series of elements \(z_{t}\) one starts by defining a objective function. The objective function is a sum of cross-correlation functions which capture the autocorrelation of the returns as well as the leverage effect and volatility clustering. The autocorrelation function is defined as

\begin{equation}
C_{f,g}(\tau) = \frac{<f(u_{t})g(u_{t-\tau})>}{\sqrt{<f^{2}(u_t)>}\sqrt{<g^{2}(u_t)>}} \label{eq:a1}
\end{equation}

where \(f(u)\) and \(g(u)\) are arbitrary functions of variable \(u\) and \(<>\) stands for time series averages. For example, if \(f(u)=g(u)=u-<u>\), \(C_{u-<u>,u-<u>}(\tau)\) is the usual textbook lag \(\tau\) autocorrelation function.

In order to keep the notation simple and without loss of generality, the returns \(x_t\) have the mean removed before calculating \(C\) in Equation \eqref{eq:a1}. Therefore the quantity of interest here is

\begin{equation}
\rho(x) = \sum_{\tau=1}^{L} C_{x,x}(\tau)+C_{x,|x|}(\tau)+\sum_{\tau=1}^{K} C_{|x|,|x|}(\tau)+C_{x^2,x^2}(\tau)  \label{eq:a2}
\end{equation}

where the first term accounts for the autocorrelation of the returns, the second term accounts for the leverage effect, the last 2 terms for the volatility clustering and \(L\) and \(K\) is the maximum lag \(\tau\) included in the sum. Finally the objective function \(\Delta\) is a function of Equation \eqref{eq:a2}

\begin{equation}
\Delta = |\rho(x)-\rho(z)| \label{eq:o1}
\end{equation}

where \(x\) is the time series of the actual data and \(z\) is the time series of the Monte Carlo surrogates. Equation \eqref{eq:o1} is minimized using simulated annealing proposed in \cite{Schreiber1998}. The simulated annealing algorithm is implemented by the nonlinear time series analysis software TISEAN \cite{tisean} using its default parameters. A simple sample code written in Julia that illustrates the algorithm for the S\&P 500 data can be found on github \footnote{https://github.com/silvaac/SMCarticle}.

\begin{figure}
\centering
\includegraphics{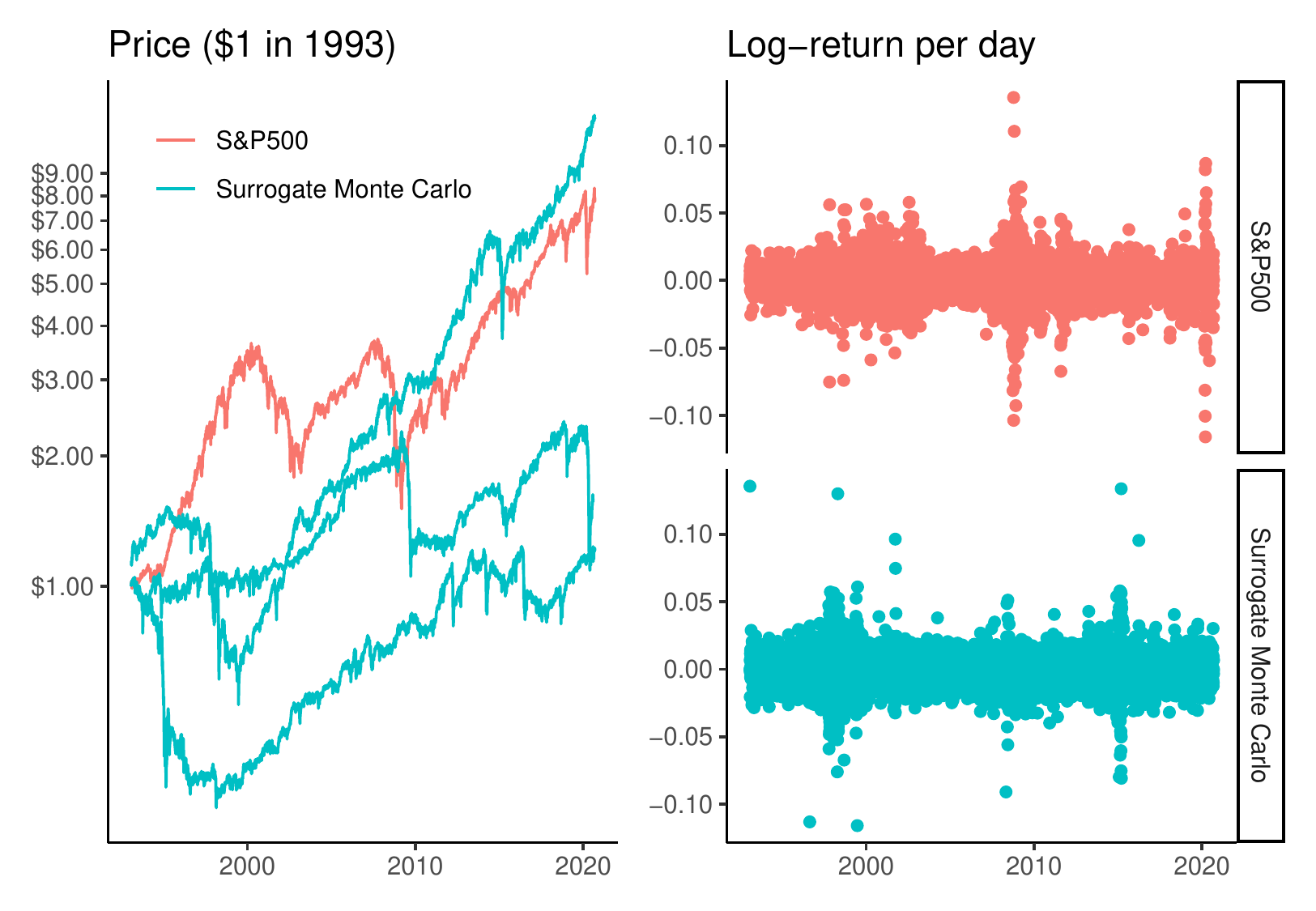}
\caption{\label{fig:vvFig}Left Panel: Time series of the price of the S\&P 500 together with three realization of the surrogate monte carlo generated data. Right panel: daily log-return of artificial time series as well as the actual S\&P 500 data from 1993-02 to 2020-08.}
\end{figure}

Figure \ref{fig:vvFig} compares the S\&P 500 time series with the surrogate Monte Carlo generated time series after 100 million optimization steps. Surrogate Monte Carlo (SMC) produces data which is visually similar to the original time series. Notice that SMC log-returns show bursts of large and persistent returns as well as occasional sharp drops similar to the actual returns.

\begin{figure}
\centering
\includegraphics{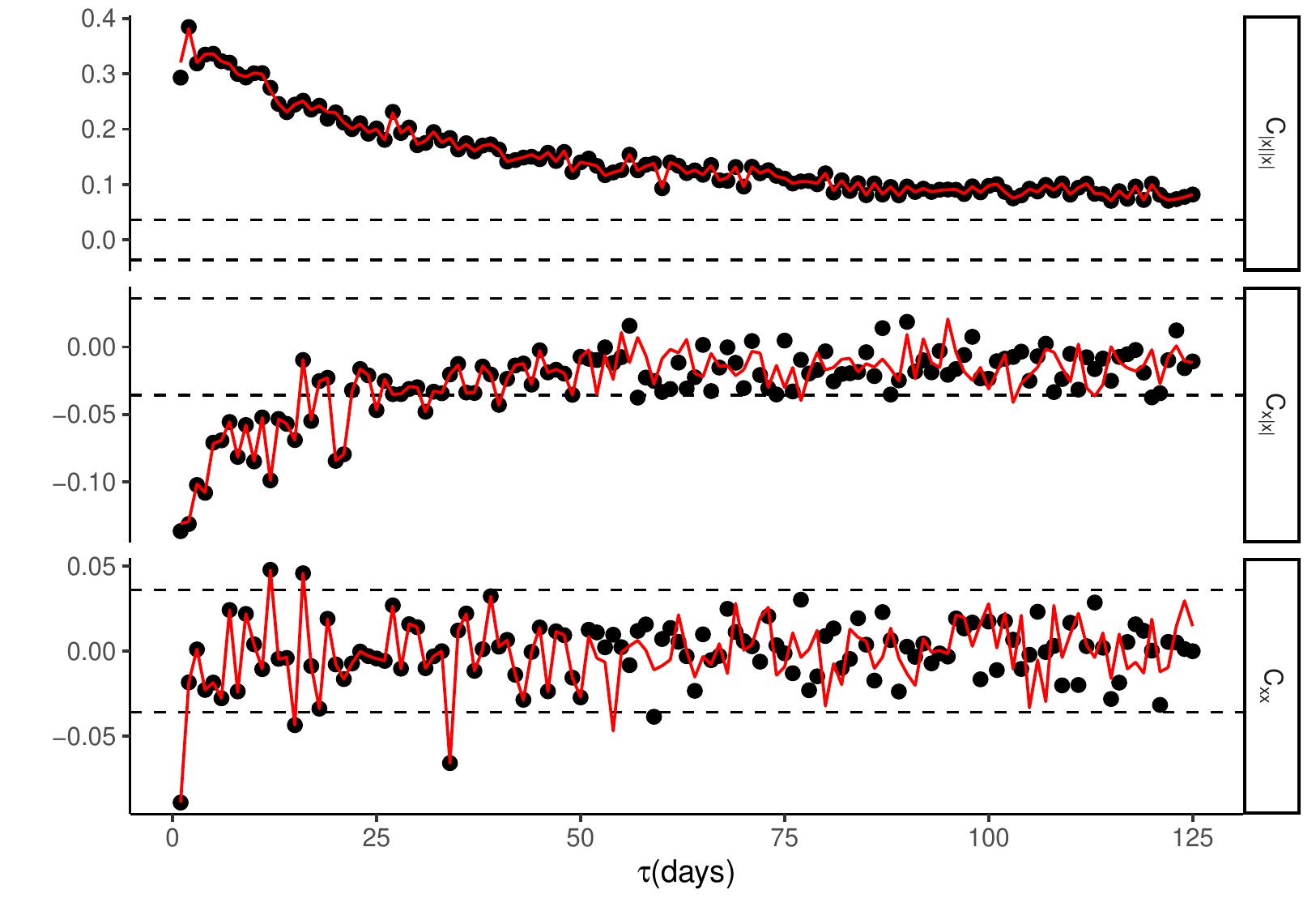}
\caption{\label{fig:vvFig2}Autocorrelation as function of time lag in days. Symbols are for the S\&P 500 and the solid red lines come from the surrogate Monte Carlo algorithm. Dashed horizontal lines: 99\% confidence level. Top: autocorrelation of the daily absolute returns ($ C_{|x|,|x|}(\tau)$). Middle: crosscorrelation of the daily returns with the daily absolute returns ($ C_{x,|x|}(\tau)$). Bottom: autocorrelation of the daily returns($C_{x,x}(\tau)$).}
\end{figure}

Figure \ref{fig:vvFig2} illustrates the convergence of the optimization algorithm using \(L=40\) and \(K=200\) in Equation \eqref{eq:a1}. The choice of values for L and K was sufficient for convergence within the 99\% confidence level. The solid red line is SMC generated data and the black circles the actual data. The agreement is very good.

\hypertarget{conclusion}{%
\section{Conclusion}\label{conclusion}}

Surrogate Monte Carlo algorithm builds on the idea that simulated data is an approximation of the actual data. Therefore if one knows the most important features that describe the data well, one can produce data that complies with these features using the simple algorithm described here. This approach is different from the traditional use of surrogates in hypothesis testing where one tries to isolate a feature to test the significance of an other \cite{LANCASTER20181}.

The complexity of this algorithm is deciding on the features one wants to use as constraints as well as its computational cost. The optimization problem is NP hard with no general convergence guarantee. However, the experiments conducted in this article did show good convergence. The algorithm appears to be well suited to model stochastic processes in general and financial data in particular.

\hypertarget{appendix}{%
\section{Appendix}\label{appendix}}

\hypertarget{details-on-convergence}{%
\subsection{Details on convergence}\label{details-on-convergence}}

The importance of the probability density is illustrated in this section. Suppose one wants to exactly replicate a deterministic function by reordering random numbers. This is precisely surrogate monte carlo (SMC) where the objective function is the deterministic function itself. The quality of the replication depends only on the random number sample. First one needs to draw random numbers which conform with the probability density of the actual process. Second, one needs to have enough samples.

\begin{figure}
\centering
\includegraphics{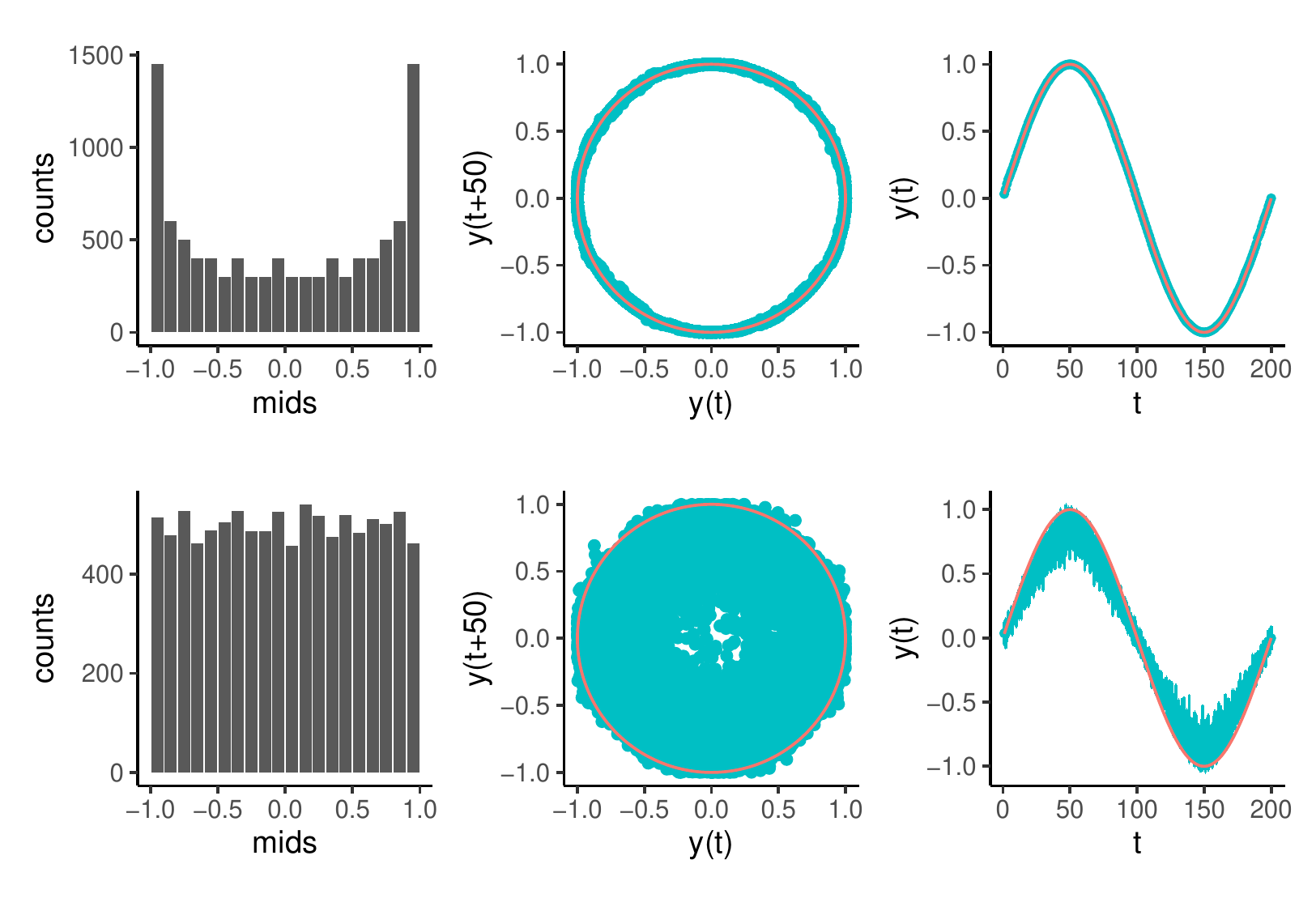}
\caption{\label{fig:sinFig}Top row: histogram for the sine function from which SMC draws random numbers. To the right the phase space diagram of the sine function (solid line) and the recovered diagram (points) by applying SMC to the random numbers. Finally, top right shows the one period of the sine function together with the average plus/minus one standard deviation over 50 periods of the SMC reconstructed sine. Bottom row: the histogram of uniform random numbers between minus one and positive one and the resulting phase diagram followed by one period time series. The difference between the rows illustrates the importance of using the correct probability density.}
\end{figure}

Take the following deterministic function \(y(t) = \sin(2\pi t/T)\) with a period of \(T=200\). Figure \ref{fig:sinFig} illustrates the importance of the using the ``correct'' unconditional probability density by comparing the empirical probability density build looking at the sinusoidal time series of 10000 data points and the uniform probability density between minus one and one. 

The experiment is as follows, first draw 10000 random numbers using either probability distribution (PDF illustrated in the first column of Figure \ref{fig:sinFig}). Next apply random permutations to these random numbers per SMC algorithm with the goal of replicating original sinusoidal. Columns 2 and 3 of Figure \ref{fig:sinFig} shows the quality of the agreement by stopping the algorithm after 100 million steps. In particular, the phase diagram (middle column) is a circle of radius one as theoretically expected where as if one starts with uniform random numbers one recovers a nearly solid disk. 

\begin{figure}
\centering
\includegraphics{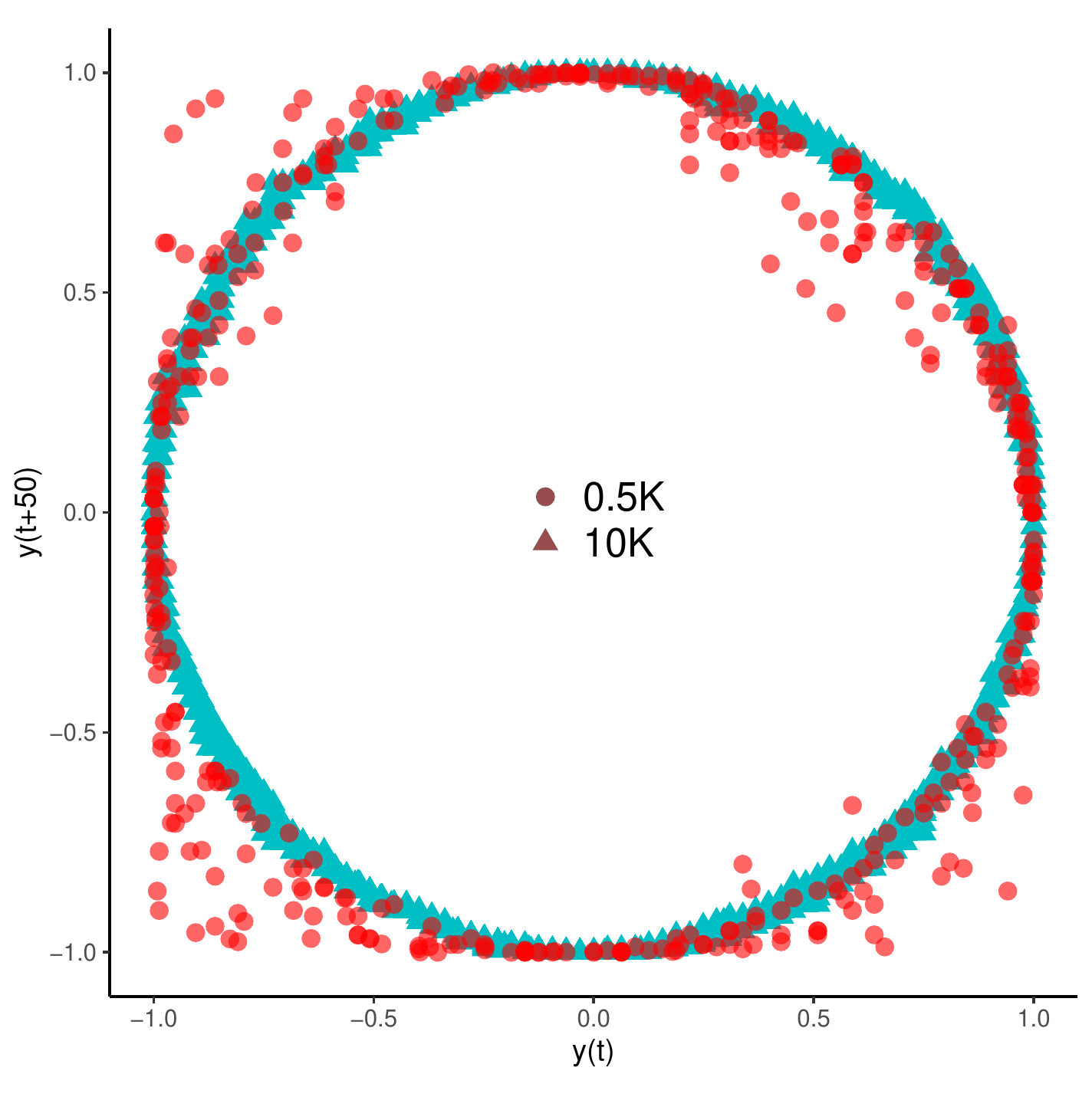}
\caption{\label{fig:sinFig2} Phase diagram of SMC reconstructed sinusoidal starting with 500 (0.5K) data points (red circles) compared to 10000 (10K) data points (blue triangles). The quality of the recovered sinusoidal using 500 data points is much worse (red circles vs blue triangles).}
\end{figure}

The effect of the length of the time series can be illustrated by comparing a time series of 500 points with the original time series of 10000 points for the same sinusoidal. Less samples clearly affects the capacity of empirically estimating the probability density and therefore the simulated time series is a worse approximation of the original function. Figure \ref{fig:sinFig2} shows that the recovered sinusoidal (red circles) shows gaps and its phase diagram is very noisy if compared to the expected unit circle.

\hypertarget{toy-example}{%
\subsection{Toy example}\label{toy-example}}

\begin{figure}
\centering
\includegraphics{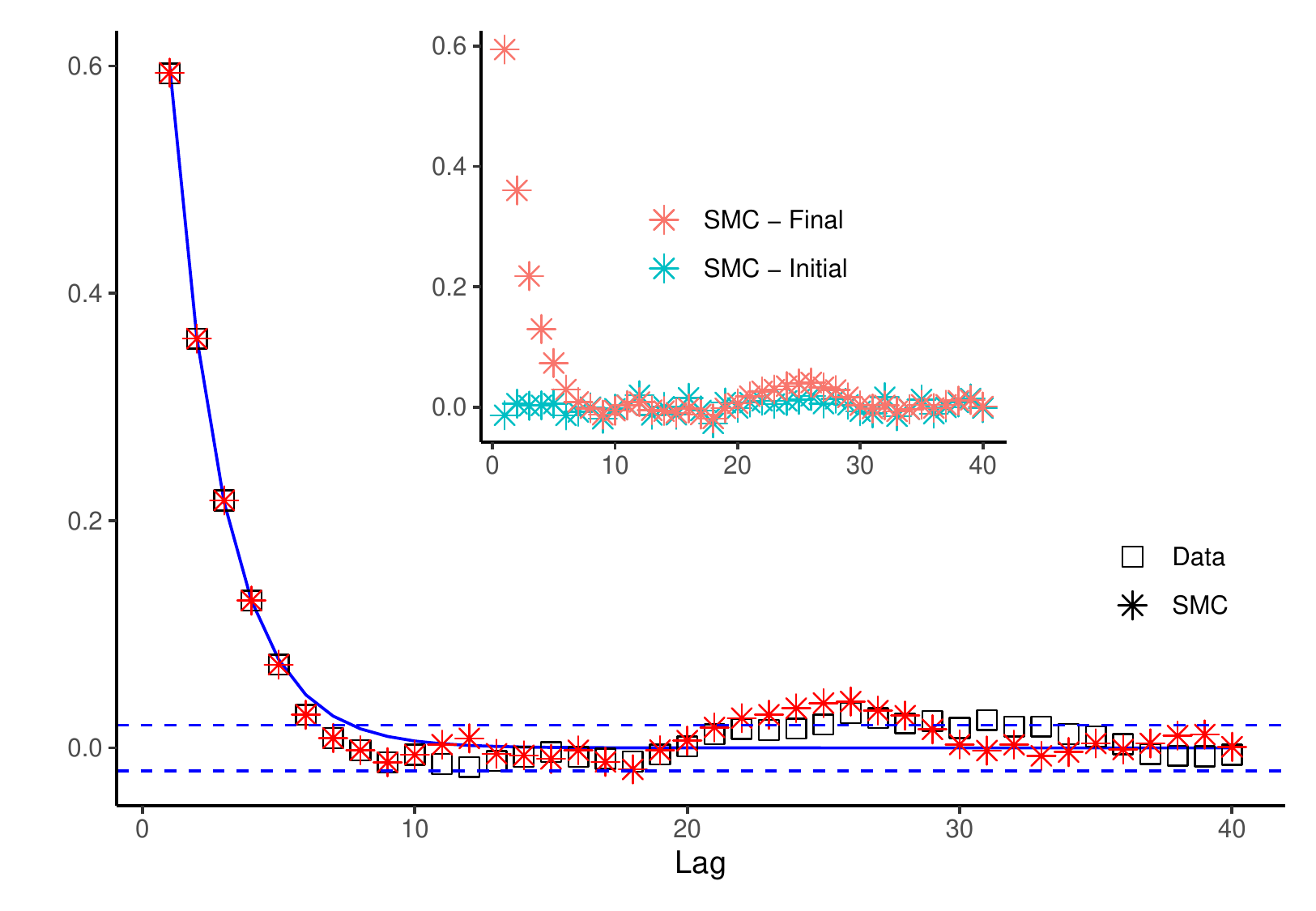}
\caption{\label{fig:unnamed-chunk-2}Autocorrelation function of an AR(1) process for 10000 data points (symbols) as well as the theoretical curve (solid blue line). Surrogate monte carlo (SMC) algorithm starts from independent Gaussian random variables and converges to the empirical autocorrelation of the AR(1) process after approximately 600 thousand iterations.}
\end{figure}

We simulate an \(AR(1)(p=0.6)\) time series with coefficient \(p=0.6\) and compare to the SMC generated time series in order to test the SMC algorithm in a controlled manner. The objective function here is the autocorrelation function calculated up to 10 lags. Figure (\ref{fig:unnamed-chunk-2}) shows the autocorrelation before and after optimization as well as the quality of the convergence. SMC starts with uncorrelated Gaussian random numbers and builds correlated random numbers that agree with the empirical autocorrelation of the \(AR(1)(p=0.6)\) process.

One can also get an idea on the rate of convergence of the surrogate monte carlo (SMC) algorithm by applying SMC to \(AR(1)\) process with different autocorrelation coefficients. It is expected that larger autocorrelation will lead to a longer run time since one starts with a time series of white noise and then recovers the autocorrelation by re-arranging the order of elements. Therefore, larger autocorrelation reduces the possible permutations of the original time series that conform with the constraints. The SMC algorithm take approximately \(3.5\) times more time to converge for a \(p=0.8\) (2.1 million iterations) than for a \(p=0.6\) (0.6 million iterations).

\section{Acknowledgments}

ACS thanks Andrei Da Silva for his help in writing the sample Julia code.

\bibliography{biblio}

\end{document}